\RequirePackage[l2tabu, orthodox]{nag}
\documentclass[dvipsnames,sigplan,10pt,authorversion]{acmart}
\renewcommand\footnotetextcopyrightpermission[1]{}
\startPage{1}

\setcopyright{none}

\usepackage[utf8]{inputenc}

\usepackage{xcolor}
\usepackage{xspace}

\usepackage{booktabs}   %% For formal tables:
\usepackage{subcaption} %% For complex figures with subfigures/subcaptions
\usepackage{mystyle}
\usepackage{mathtools}

\usepackage{cmds}

\begin{document}

%% Title information
\title{Translation Validation for Security Properties}         %% [Short Title] is optional;
                                        %% when present, will be used in
                                        %% header instead of Full Title.
%\titlenote{with title note}             %% \titlenote is optional;
                                        %% can be repeated if necessary;
                                        %% contents suppressed with 'anonymous'
\subtitle{(Extended Abstract)}                     %% \subtitle is optional
%\subtitlenote{with subtitle note}       %% \subtitlenote is optional;
                                        %% can be repeated if necessary;
                                        %% contents suppressed with 'anonymous'

%% Author information
%% Contents and number of authors suppressed with 'anonymous'.
%% Each author should be introduced by \author, followed by
%% \authornote (optional), \orcid (optional), \affiliation, and
%% \email.
%% An author may have multiple affiliations and/or emails; repeat the
%% appropriate command.
%% Many elements are not rendered, but should be provided for metadata
%% extraction tools.

%% Author with single affiliation.
\author{Matteo Busi}
%\authornote{with author1 note}          %% \authornote is optional;
                                        %% can be repeated if necessary
%\orcid{nnnn-nnnn-nnnn-nnnn}             %% \orcid is optional
\affiliation{
  % \position{Position1}
  %\department{Dipartimento di Informatica}              %% \department is recommended
%
    \institution{Universit\`a di Pisa}            %% \institution is required
%  \institution{Universit\`a di Pisa, Pisa, Italy}            %% \institution is required
  %
  %\streetaddress{Street1 Address1}
 \city{Pisa}
  %\state{State1}
  %\postcode{Post-Code1}
  \country{Italy}                    %% \country is recommended
}
\email{matteo.busi@di.unipi.it}          %% \email is recommended

%% Author with single affiliation.
\author{Pierpaolo Degano}
%\authornote{with author1 note}          %% \authornote is optional;
                                        %% can be repeated if necessary
%\orcid{nnnn-nnnn-nnnn-nnnn}             %% \orcid is optional
\affiliation{
  % \position{Position1}
  %\department{Dipartimento di Informatica}              %% \department is recommended
%  \institution{Universit\`a di Pisa, Pisa, Italy}            %% \institution is required
%
  \institution{Universit\`a di Pisa}            %% \institution is required
%
  %\streetaddress{Street1 Address1}
 \city{Pisa}
  %\state{State1}
  %\postcode{Post-Code1}
 \country{Italy}                    %% \country is recommended
}
\email{degano@di.unipi.it}          %% \email is recommended

%% Author with single affiliation.
\author{Letterio Galletta}
%\authornote{with author1 note}          %% \authornote is optional;
                                        %% can be repeated if necessary
%\orcid{nnnn-nnnn-nnnn-nnnn}             %% \orcid is optional
\affiliation{
  % \position{Position1}
  %\department{Dipartimento di Informatica}              %% \department is recommended
  %
   \institution{\mbox{IMT School for Advanced Studies}}%% \institution is required
   %
%  \institution{\mbox{IMT School for Advanced Studies, Lucca, Italy}}%% \institution is required
  %\streetaddress{Street1 Address1}
 \city{Lucca}
  %\state{State1}
  %\postcode{Post-Code1}
 \country{Italy}                    %% \country is recommended
}
\email{letterio.galletta@imtlucca.it}          %% \email is recommended

\begin{abstract}
  Secure compilation aims to build compilation chains that preserve security properties when translating programs from a source to a target language.
  Recent research led to the definition of secure compilation principles that, if met, guarantee that the compilation chain in hand never violates specific families of security properties.
  Still, to the best of our knowledge, no effective procedure is available to check if a compilation chain meets such requirements.
  Here, we outline our ongoing research inspired by translation validation, to effectively check one of those principles.
\end{abstract}


\maketitle

\section{Introduction} \label{sec:introduction}
% !TeX root = ../prisc2019_main.tex

\emph{Secure compilation} is concerned with ensuring that the security properties at the source level are preserved as they are at the target level or, equivalently, that all the attacks that can be carried out at the target level are also possible at the source level. In this way, it is enough to reason at the source level to rule out attacks at all.

Consider a functional and reactive source language with I/O \emph{primitives}, but none for communication,
and a compiler to a target language that relies on system calls for managing the I/O (on screen, network, etc.).
A run of the compiler transforms the source program
\begin{align*}
\S \triangleq \src{\ffun{i}{ \fifthenelse{i \geq 0}{(\fseq{\fapp{\mathbf{print}}{i}}{i})}{(-1)}}}
\end{align*}
into the target program
\begin{align*}
\T \triangleq \trg{\ffun{i}{ \fifthenelse{i \geq 0}{(\fseq{\fapp{\mathtt{sc\_print}}{i}}{i})}{(-1)}}}
\end{align*}
(we highlight in $\src{\text{blue}}$ the elements of the source language, and in
$\trg{\text{red}}$ the elements of the target language for better readability).
Although correct,
this compilation does not preserve the security property requiring a program to
never send a value on the network, which $\S$ enjoys in \emph{any} context -- an expression with a single hole.
This property still holds when $\T$ is plugged into a \emph{non-evil} target context that correctly implements the system calls.
Instead things go wrong when the context is \emph{evil}, i.e.\ it maliciously implements the system calls.
For example, the property is violated when we plug $\T$ into
\begin{align*}
    \ctxtn{C}{T}{\mathit{evil}} \triangleq \trg{\fapp{
        \big(\ffun{i}{
            \flet{\mathtt{sc\_print}}
                {
                    \ffun{x}{
                        (\fseq
                            {\fapp{\mathbf{display}}{x}}
                            {\fapp{\mathbf{send}}{x}})
                        }
                }
                {\fapp{\hole{\cdot}}{i}}
        }\big)
    }{42}}.
\end{align*}

Our idea is to provide a method, inspired by {\em translation validation (TV)}~\cite{pnueli1998translation}, that we call {\em secure translation validation (STV)}.
It {\em automatically} decides if a compiler preserves a family of hyperproperties of interest, for a given program $\Prg$.
%.
STV is carried on at {\em load time} and we argue that this is the right time.
On the one hand, it is not too early because one typically wants some security guarantees on a module, e.g.\ a library, before launching a program using it.
On the other hand, it is not too late, since executing the same program in different contexts results in different security guarantees.

%%% Local Variables:
%%% mode: latex
%%% TeX-master: "../prisc2019_main"
%%% End:

\section{Our proposal} \label{sec:proposal}
% !TeX root = ../prisc2019_main.tex
The technique TV checks the correctness of the compilation of a given program $\Prg$, rather than proving the compiler correct for all inputs.
Roughly, it works as follows:
first, the source and the target languages are endowed with semantics sharing the same observables; then a suitable simulation is defined between the result of the compilation and the corresponding source program: if such a simulation exists, the compiler is correct;
finally, an algorithm effectively computes the required simulation, if any.
Remarkably, this algorithm gives a fully automatic way of checking the correctness of real compilers~\cite{necula2000translation}.
A tempting approach could be mechanically proving also the security of a compiler by showing (the existence of) a (suitable) simulation between the source and the target program.
However, the construction of the required simulation, if any, is undecidable when the program in hand is not finite-state~\cite{deng2016securing}.
Static analysis comes to our rescue and
allows us to devise a mechanical (and approximated) procedure to deal with this
problem. 

More precisely, we proceed as follows.
At load time we plug the compiled program $\compnolang{P}$ into the (target) context, obtaining $\ctxhtn{C}{T}{}{\compnolang{\Prg}}$,
the behaviour of which is safely over-approximated by a static analysis.
An approximation is a \emph{history expression}~\cite{bartoletti2009local}, i.e.\ a (finite-state) process of a basic process algebra~\cite{bergstra1985algebra}, whose actions are the observables of
the {\em trace semantics} of the target and source languages.
For example, in the code above the observable of the primitive {\bf print} will be {\bf display}.

Once the history expression for $\ctxhtn{C}{T}{}{\compnolang{\Prg}}$ is computed, we verify on it if the compilation process broke some of the properties of interest.
The actual verification depends on the family of properties we are interested in.
A first principle that one might consider is {\em full abstraction (FA)}~\cite{abadi1999protection}:
\begin{align*}
    \forall \src{P_1}, \src{P_2} \,.\, \src{P_1} \mathrel{\src{\simeq}} \src{P_2} \Leftrightarrow \compnolang{P_1} \mathrel{\trg{\simeq}} \compnolang{P_2}
\end{align*}
where $\src{P_1}$ and $\src{P_2}$ are programs, while $\src{\simeq}$ and $\trg{\simeq}$ are suitable notions of \emph{behavioural equivalence}.
However, FA has well-known shortcomings~\cite{patrignani2017secure} and does not fit well with STV because of the universal quantification over pairs of programs.

Actually, the principles proposed by Abate et al.~\cite{abate2018exploring} are
more appropriate for STV purposes, in particular \emph{robustly safe compilation (RSC)}:
\begin{align*}
    \forall \Prg, \ctxtn{C}{T}{}, m. \, \big(
     m \in \pref{\ctxhtn{C}{T}{}{\compnolang{\Prg}}}
        \Rightarrow
        \exists \ctxsn{C}{S}{}.\, m \in \pref{\ctxhsn{C}{S}{}{\Prg}}
                             \big).
\end{align*}
Intuitively, RSC considers finite traces produced by the compiled program when plugged in a possibly evil context: the compiler preserves all the safety properties \emph{iff} there exists a context in which the source program also produces the same finite trace.
The operator $\pref{\cdot}$ returns the set of the prefixes of the traces of its argument.

We can effectively check this principle by using STV.
Indeed, we can get rid of the universal quantifiers on programs and contexts because STV only considers a single program at a time and is performed at load time.
Given a program $\Prg$ and a context $\ctxtn{C}{T}{}$, it suffices then verifying the following
\begin{align*}
    \mathit{STV}_{\mathit{RSC}} \triangleq
        m \in \pref{\trg{H_T}\xspace}
            \Rightarrow
        \exists \ctxsn{C}{S}{}.\, m \in \pref{\src{H_S}\xspace}
\end{align*}
where $\trg{H_T}\xspace$ is the history expression associated with $\ctxhtn{C}{T}{}{\compnolang{\Prg}}$ and $\src{H_S}\xspace$ that associated with $\ctxhsn{C}{S}{}{\Prg}$.

Since history expressions \emph{safely} approximate the behaviour of programs,
their semantics includes the set of traces of the program they are associated with.
Also, since the properties of interest are defined in terms of traces,
STV succeeds when a $\ctxsn{C}{S}{}$ with the desired property can be proved to exist starting from $\trg{H_T}\xspace$.
Note that, since history expressions are processes of a basic process algebra, it is decidable whether a prefix belongs to the semantics of a history expression.
However, there is a price to pay in order to have an effective procedure.
False negatives may be produced, and we may fail to prove a compilation secure because history expressions \emph{over}-approximate the behaviour of programs.

To intuitively illustrate the idea, recall the example above and assume to design our history expressions to track the I/O actions of a program.
Consider now the history expression associated with $\T$ plugged into the evil context $\ctxtn{C}{T}{\mathit{evil}}$:
\begin{align*}
    \trg{H_{\mathit{evil}}} = \trg{(\mathbf{display}\cdot \mathbf{send}) + \epsilon}
\end{align*}
Intuitively it represents that $\ctxhtn{C}{T}{}{\T}$ writes on the screen and sends something on the network, or does nothing ($\trg{\epsilon}$).
The prefix $\trg{\mathbf{display} \cdot \mathbf{send}}$ of $\trg{H_T}$ has no counterpart in any source level context (recall that the source language cannot perform I/O on the network).
So, this program and context combination is rejected by our analysis.

Instead, plugging $\T$ into the following non-evil context:
\begin{align*}
    \ctxtn{C}{T}{\mathit{friendly}} \triangleq \trg{\fapp{
            (\ffun{i}{
                \flet{\mathtt{sc\_out}}
                    {
                        \ffun{x}{
                            \fapp{\mathbf{display}}{x}}
                    }
                    {\fapp{[\cdot]}{i}}
            })
        }{42}}
\end{align*}
results in the history expression
\begin{align*}
    \trg{H_{\mathit{friendly}}}\xspace = \trg{\mathbf{display} + \epsilon}
\end{align*}
that has an acceptable counterpart in the source contexts.
Our property thus holds and security is preserved.

Our approach works also for program optimizations.
For example, consider a source program $\src{S'}\xspace$ that has a choice between two behaviour, both prefixed by an output on the screen. e.g.\ a warning to the user.
Its history expression will essentially be as follows:
\begin{align*}
    \src{H_{S'}} = \src{((\mathbf{display} \cdot H_{\mathit{1}}) + (\mathbf{display} \cdot H_{\mathit{2}}))}
\end{align*}
Any optimizing compiler will detect that both branches share the same output, and will factor it out of the choice.
The history expression associated with the optimized program $\trg{T'}$ will be
\begin{align*}
    \trg{H_{T'}} = \trg{\mathbf{display} \cdot ( H_{\mathit{1}} + H_{\mathit{2}})}
\end{align*}
Plugging $\trg{T'}$ into a non-evil target context results in $\ctxhtn{C'}{T}{}{\trg{T'}}$ (with history  expression $\trg{H_{\ctxhtn{C'}{T}{}{\trg{T'}\xspace}}}$).
It is not difficult finding a source level context $\ctxsn{C'}{S}{}$ such that any prefix of $\ctxhtn{C'}{T}{}{\trg{T'}}$ has a counterpart in $\ctxhsn{C'}{S}{}{\src{S'}\xspace}$ (with history expression $\src{H'_{\ctxhsn{C'}{S}{}{\src{S'}\xspace}}}$).
The task is easy, because $\trg{H_{\ctxhtn{C'}{T}{}{\trg{T'}\xspace}}}$ and $\src{H'_{\ctxhsn{C'}{S}{}{\src{S'}\xspace}}}$ have the same semantics --- in this case equivalence is trivial, while in more complex cases one can use the equational theory over history expressions, which is decidable~\cite{bartoletti2009local}.

We briefly discussed examples showing how safety property preservation can be effectively checked.
We are currently extending STV to deal with safety hyperproperties, and we are confident that also other families of properties can fit our proposal.

%%% Local Variables:
%%% mode: latex
%%% TeX-master: "../prisc2019_main"
%%% End:

%% Bibliography
\bibliography{biblio}

\end{document}